\documentstyle[12pt,graphicx,caption2]{article}

\begin{document}

\title{$n\bar{n}$ conversion in finite nuclei}
\author{V.I.Nazaruk\\
Institute for Nuclear Research of RAS, 60th October\\
Anniversary Prospect 7a, 117312 Moscow, Russia.}

\date{}
\maketitle
\bigskip

\begin{abstract}
The new model of $n\bar{n}$ transitions in nuclei based on unitary $S$-matrix is
considered. The $|in>$-state of nucleus is described by single-particle shell model. 
The dynamical process part is calculated by means of field-theoretical approach with 
finite time interval. The lower limit on the free-space $n\bar{n}$ oscillation time 
$\tau _{{\rm min}}$ is in the range $10^{16}\; {\rm yr}>\tau _{{\rm min}}>1.2\cdot 10^{9}\; {\rm s}$.
\end{abstract}

\vspace{5mm}
{\bf PACS:} 11.30.Fs; 13.75.Cs

\vspace{5mm}
Keywords: diagram technique, infrared divergence, single-particle shell model

\vspace{1cm}

*E-mail: nazaruk@inr.ru

\newpage
\setcounter{equation}{0}
\section{Introduction}
Any information on the occurrence of $n\bar{n}$ oscillation [1,2] is important in order to
discriminate among various grand unified theories. The most direct limit on the free-space
$n\bar{n}$ oscillation time $\tau _{{\rm min}}$ is obtained using free neutrons: $\tau _{{\rm min}}=
0.86\cdot 10^{8}\; {\rm s}$ [3]. Alternatively, a limit can be extracted from the nuclear
annihilation lifetime measured in proton-decay type experiments (see, for example, Refs. [4-11]). In 
this case one should calculate the $n\bar{n}$ transition in nuclei followed by annihilation:
\begin{equation}
(\mbox{nucleus})\rightarrow (\bar{n}-\mbox{nucleus})\rightarrow M,
\end{equation}
where $M$ are the annihilation mesons. The analogous process in the medium is
\begin{equation}
(n-\mbox{medium})\rightarrow (\bar{n}-\mbox{medium})\rightarrow M.
\end{equation}
The particle oscillations in absorbing matter take place.

In the standard calculations of $ab$ oscillations in the medium [12-14] the interaction of 
particles $a$ and $b$ with the matter is described by the potentials $U_{a,b}$ (potential 
model). ${\rm Im}U_b$ is responsible for loss of $b$-particle intensity. In particular, 
this model is used for the processes (1) and (2) [4-11]. 

In [10,11] it was shown that one-particle (potential) model mentioned above does not describe the
processes (1) and (2) and thus total neutron-antineutron transition probability. For instance, 
the total neutron-antineutron transition probability given by the potential 
model is $W\sim 1/\Gamma $ ($\Gamma $ is the annihilation width of $\bar{n}$ in the medium),
whereas the realistic calculation gives $W\sim \Gamma $ (see Sect. 5). In the potential model
the effect of final state absorption (annihilation) acts in the opposite (wrong) direction, which 
tends to the additional suppression of the $n\bar{n}$ transition. So the potential model should be 
rejected. The $S$-matrix should be {\em unitary}.
 
(For the oscillations in the external field [15,16] the Hamiltonian is hermitian and so the 
absorption is described correctly. The above-given remark holds only for the processes (1) and (2) 
calculated by means of potential model (non-hermitian Hamiltonian). We also note that the potential 
model describes correctly the channel with $\bar{n}$ in the final state [11].)

The unitarity of the $S$-matrix means that new model should be developed. In [8,9] we have proposed the 
model of the $n\bar{n}$ transition in medium followed by annihilation which does not contain the 
non-hermitian operators. It is shown in Fig. 1a. The results are summarized and discussed in [17]. 
In the present paper the $n\bar{n}$ transitions in finite nuclei followed by annihilation (process 
(1)) are considered. The process model is shown in Fig. 1b. The reason is that the limit is extracted 
from the nuclear annihilation lifetime and so one should calculate the process (1) and not (2). As we 
shall see later, the results are the same as for nuclear matter. However, this fact is not obvious
since the calculations for the processes (1) and (2) are essentially different. A distinguishing
feature of the problem under study is the zero momentum transfer in the $n\bar{n}$ transition vertex.
Because of this the $S$-matrix amplitudes corresponding to the processes shown in Figs. 1a and 1b 
contain infrared divergence. The problem of infrared divergence for the particle in the bound state 
(Fig. 1b) is considered for the first time. To gain a better understanding of the material, it is 
desirable to look through the Ref. [17].

The paper is organized as follows. In Sect. 2 we formulate the models for the processes
(1) and (2). In Sect. 3 the diagram 1b corresponding to the model with bare propagator is 
calculated. In Sect. 4 it is shown that in the case of $S$-matrix problem formulation the 
process amplitude corresponding to the model with bare propagator is singular. The model 
with the dressed propagator is studied in Sect. 5. The results are summarized and discussed 
in Sect. 6.

\section{Models}
The qualitative process picture is as follows. The free-space $n\bar{n}$ transition comes 
from the exchange of Higgs bosons with the mass $m_H>10^5$ GeV [2] and so the subprocess
of $n\bar{n}$ conversion is scarcely affected by a medium effects. From the dynamical point 
of view this is a momentary process: $\tau _c\sim 1/m_H<10^{-29}$ s. The antineutron 
annihilates in a time $\tau _a\sim 1/\Gamma $. We deal with two-step process with the 
characteristic time $\tau _2\sim \tau _a$.

Thus, the localization of the neutron incide the nucleus {\em does not tend to suppress} of 
the $n\bar{n}$ conversion. This can be also understood using the analogy with the nuclear
$\beta $ decay and decay of free neutron. It should be emphasized that above-given
qualitative process picture {\em does not contradict} to well-known results on particle 
oscillations except the absorption channel (see Sect. 5.2 of Ref. [9]).

We consider Fig. 1a. If the antneutron propagator is bare, it contains the infrared 
singularity conditioned by zero momentum transfer in the $n\bar{n}$ transition vertex. This 
circumstance changes the standard calculation scheme radically. The same is true for the 
Fig. 1b (see Sect. 4). Since the process (2) has been considered in details [9,17], we 
draw analogy with the model used for the diagram 1a.

We return to Fig. 1a. The neutron potential $U_n$ is included in the neutron wave 
function (unperturbed Hamiltonian): 
\begin{equation}
n(x)=\Omega ^{-1/2}\exp (-ipx).
\end{equation}
Here $p=(\epsilon ,{\bf p})$ is the neutron 4-momentum; $\epsilon ={\bf p}^2/2m+U_n$.
The interaction Hamiltonian is
\begin{eqnarray}
{\cal H}_I={\cal H}_{n\bar{n}}+{\cal H},\nonumber\\
{\cal H}_{n\bar{n}}=\epsilon _{n\bar{n}}\bar{\Psi }_{\bar{n}}\Psi _n+H.c.,\nonumber\\
H_I(t)=\int d^3x{\cal H}_I(x).
\end{eqnarray}
Here ${\cal H}_{n\bar{n}}$ and ${\cal H}$ are the Hamiltonians of the $n\bar{n}$ transition 
[4] and the $\bar{n}$-medium interaction, respectively; $\epsilon _{n\bar{n}}$ is a small 
parameter with $\epsilon _{n\bar{n}}=1/\tau $, where $\tau $ is the free-space $n\bar{n}$ 
oscillation time; $\Psi _n$ and $\Psi _{\bar{n}}$ are the operators of the neutron and
antineutron fields; $m_n=m_{\bar{n}}=m$. 

\begin{figure}[h]
  {\includegraphics[height=.25\textheight]{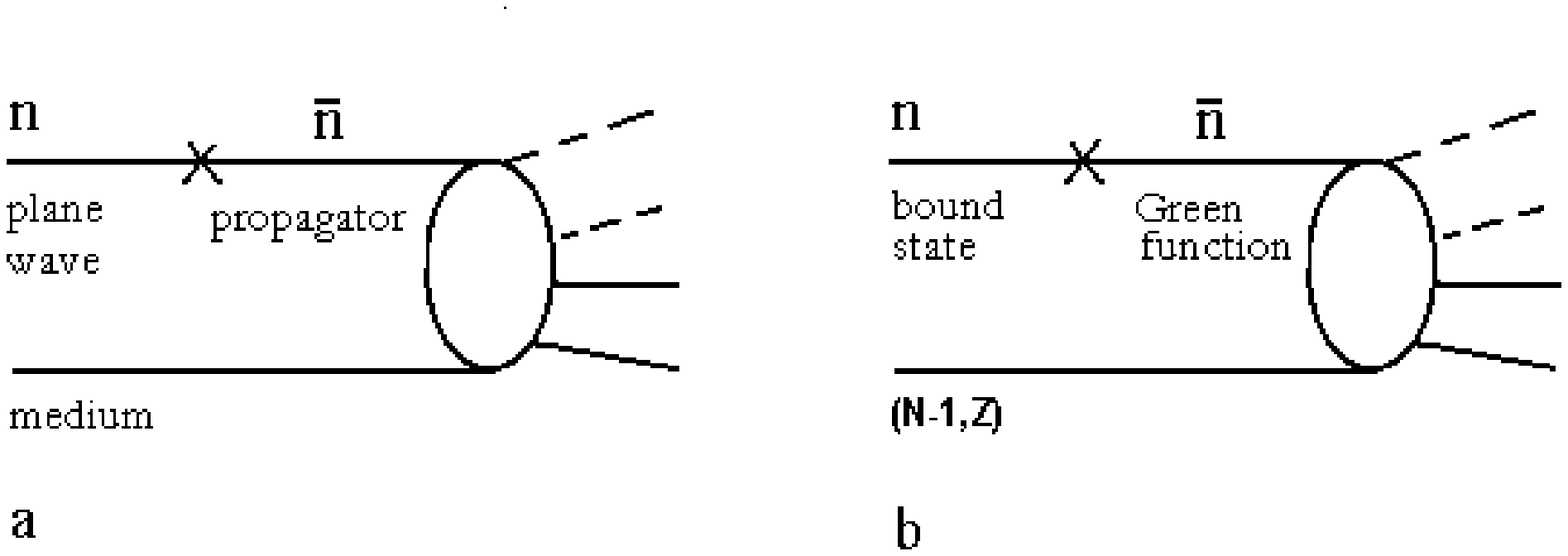}}
  \caption{$n\bar{n}$ transition in the medium ({\bf a}) and nuclei ({\bf b}) followed 
by annihilation.}
\end{figure}

For the process shown in Fig. 1b we take the single-particle shell model of nucleus. The 
initial neutron state is defined by equation of motion:
\begin{eqnarray}
(i\partial_t-H_0)n(x)=0,\nonumber\\
H_0=-\nabla^2/2m+U,
\end{eqnarray}
where $U$ is the self-consistent neutron potential. The interaction Hamiltonian is given 
by (4), where ${\cal H}$ is the Hamiltonian of the $\bar{n}$-nuclear interaction.

The neutron state is stationary:
\begin{equation}
n_j(t,{\bf x})=e^{-i\epsilon _jt}n_j({\bf x}),
\end{equation}
$x=(t,{\bf x})$. Here $n_j({\bf x})$ and $\epsilon _j$ are the eigenfunctions and 
eigenvalues of the Hamiltonian $H_0$:
\begin{equation}
H_0n_j({\bf x})=\epsilon _jn_j({\bf x}).
\end{equation}
The eigenfunctions $n_j({\bf x})$ form the complete orthogonal set.

The Green function of Eq. (5) is defined as
\begin{equation}
[i\frac{\partial}{\partial t'}-H_0({\bf x}')]G(x',x)=\delta ^3({\bf x}'-{\bf x}) 
\delta (t'-t).
\end{equation}

Comparing with Fig. 1a, we see that the neutron plane wave is replaced by the bound 
state wave function (6); the antineutron propagator should be replaced by the Green function $G(x',x)$ defined by (8). Both of these processes are described by identical models: The $|in>$-states are the eigenfunctions of unperturbed Hamiltonian. In the case of diagram 1a, this is the neutron plane wave. In the case of Fig. 1b, this is the bound state wave function. The interaction Hamiltonian is given by (4). This is the standard formulation of the problem which allows {\em to derive the process amplitude directly} from interaction Hamiltonian in contrast to the model based on diagram technique for direct nuclear reactions [18].

In principle, the antineutron propagator can be bare or dressed. In the latter case the
calculation is standard and simple. The diagram with bare propagator contains infrared
divergence. Since the corresponding calculations are non-typical, particular attention 
is given to the model with bare propagator.

We write the general formulas which are used below. Since $n_j(x)$ form the complete 
orthogonal set, the Green function can be represented as [19]
\begin{equation}
G(x',x)=-i\theta (t'-t)\sum_{j}n_j(x')n_j^*(x).
\end{equation}
Using the condition of completeness
\begin{equation}
\sum_{j}n_j(t,{\bf x}')n_j^*(t,{\bf x})=\delta ^3({\bf x}'-{\bf x}),
\end{equation}
one obtains the important relation
\begin{equation}
i\int d^3xG(x',x)n_j(x)=\theta (t'-t)\sum_{k}n_k(x')\int d^3xn_k^*(x)n_j(x)=\theta (t'-t)
n_j(t',{\bf x}').
\end{equation}
The RHS of Eq. (9) acts as the Feynman propagator transforming the function $n_j(x)$
into the point $x'$: $n_j(x)\rightarrow n_j(x')$. We recall that the basis of plane waves
is not used. Equations (9)-(11) are valid for any local potential $U$.

\section{Model with bare propagator}
In this section we calculate the process (1) for the model with bare propagator. The
corresponding calculations are non-trivial: the problem of infrared divergence for the 
particle in the bound state is considered for the first time. First of all we outline a 
method of calculation of the process (2) for the model with {\em bare} propagator [8,9,17] 
because for the process (1) the idea of calculation is the same. 

The amplitude corresponding to Fig. 1a diverges:
\begin{equation}
M=\epsilon _{n\bar{n}}G_0M_a,
\end{equation}
\begin{eqnarray}
G_0=\frac{1}{\epsilon _{\bar{n}}-{\bf p}_{\bar{n}}^2/2m-U_n+i0}\sim \frac{1}{0},\nonumber\\
<\!f0\!\mid T\exp (-i\int dx{\cal H}(x))-1\mid\!0\bar{n}_{p}\!>=
N(2\pi )^4\delta ^4(p_f-p_i)M_a,
\end{eqnarray}
${\bf p}_{\bar{n}}={\bf p}$, $\epsilon _{\bar{n}}=\epsilon $. Here $\mid \!0\bar{n}_p\!>$ is the state 
of the medium containing the $\bar{n}$ with the 4-momentum $p=(\epsilon ,{\bf p})$;  $N$ includes the 
normalization factors of the wave functions. $M_a$ is the antineutron annihilation amplitude. It contains 
all the $\bar{n}$-medium interactions followed by annihilation including antineutron rescattering in the 
initial state. Due to this the antineutron propagator $G_0$ is bare. 

These are infrared singularities conditioned by zero momentum transfer in the $n\bar{n}$ 
transition vertex. There is no compensation mechanism by radiative corrections. This is
unremovable peculiarity. Moreover, for the problem under study the $S$-matrix problem 
formulation $(\infty ,-\infty )$ is physically incorrect. For solving the problem the 
field-theoretical approach with finite time interval [8,20] is used. It is infrared free. 
The problem is formulated on the interval $(t/2,-t/2)$. If $H=U_{\bar{n}}=$const. 
($U_{\bar{n}}$ is the optical potential of $\bar{n}$), the approach with finite time 
interval reproduces all the well-known results on particle oscillations (see Sect. 5.2 of 
Ref. [9]).

For the process shown in Fig. 1b the zero momentum transfer also takes place and so it contains the infrared singularities as well (see Sect. 4). As with Fig. 1a, we formulate the problem on the finite time interval $(t/2,-t/2)$.  

We consider the process (1) on the finite time interval $(t/2,-t/2)$. (The case of
$S$-matrix problem formulation $(\infty,-\infty)$ is studied in next section.)
 
The vector of initial state is
\begin{equation}
\mid \!0n_j\!>=b^+_j\mid \!0\!>,
\end{equation}
where $\mid \!0n_j\!>$ is the nucleus containing the $n$ in the state $j$. Since the basis
(6) is used, in the expressions for the $\Psi $-operators $\Psi _n$ and $\Psi _{\bar{n}}$ 
the plane waves should be replaced by the eigenfunctions $n_j(x)$ (Furry representation). 
Then
\begin{equation}
\Psi _n\mid \!0n_j\!>=n_j(t,{\bf x})\mid \!0\!>,
\end{equation}
where $n_j(t,{\bf x})$ is given by (6). We introduce the evolution operator $U(t)=1+iT(t)$.
In the lowest order in $H_{n\bar{n}}$ the matrix element $T_{fi}(t)$ is
\begin{eqnarray}
T_{fi}(t)=-i<f\mid T(\exp (-i\int_{-t/2}^{t/2}dt_1H_I(t_1))-1\mid \!0n_j\!>=\nonumber\\
-\epsilon _{n\bar{n}}<f\mid \sum_{k=1}^{\infty}T_k(t)\int_{-t/2}^{t_k}dt_c\int d^3x_c
\bar{\Psi }_{\bar{n}}(x_c)\Psi _n(x_c)b^+_j\mid \!0\!>,
\end{eqnarray}
\begin{equation}
T_k(t)=(-i)^k \int_{-t/2}^{t/2}dt_1...\int_{-t/2}^{t_{k-1}}dt_kH(t_1)...H(t_k).
\end{equation}
In the last multiplier of Eq. (17) we separate out the antineutron field operator
$\Psi _{\bar{n}}(x_k)$:
\begin{equation}
H(t_k)=\int d^3x_k{\cal H}(x_k)=\int d^3x_k{\cal H}'(x_k)\Psi _{\bar{n}}(x_k).
\end{equation}
Using (15), we obtain
\begin{equation}
T_{fi}(t)=i\epsilon _{n\bar{n}}<f\mid \sum_{k=1}^{\infty}T_{k-1}(t)
\int_{-t/2}^{t_{k-1}}dt_k\int d^3x_k{\cal H}'(x_k)J_k(t_k)\mid \!0>,
\end{equation}
\begin{equation}
J_k(t_k)=\int_{-t/2}^{t_k}dt_c\int d^3x_c<T(\Psi _{\bar{n}}(x_k)
\bar{\Psi }_{\bar{n}}(x_c))>\bar{n}_j(x_c).
\end{equation}
As in the case of plane waves [17], the following relation takes place:
\begin{equation}
\int d^3x_c<T(\Psi _{\bar{n}}(x_k)\bar{\Psi }_{\bar{n}}(x_c))>\bar{n}_j(x_c)=\bar{n}_j(x_k)
\end{equation}
(Schrodinger fields). This relation is analogue of Eq. (11) in the second quantization
representation. Equation (20) becomes
\begin{equation}
J_k(t_k)=\bar{n}_j(x_k)\int_{-t/2}^{t_k}dt_c.
\end{equation}
As in (15),
\begin{equation}
\bar{n}_j(x_k)\mid \!0\!>=\Psi _{\bar{n}}(x_k)d^+_j\mid \!0\!>=\Psi _{\bar{n}}(x_k)
\mid \!0\bar{n}_j\!>,
\end{equation}
where $\mid \!0\bar{n}_j\!>$ is the $\bar{n}$-nucleus containing the $\bar{n}$ in the state
$j$ (with the energy $\epsilon _j$).
Turning back to the Hamiltonian $H(t_k)$
\begin{equation}
\int d^3x_k{\cal H}'(x_k)\Psi _{\bar{n}}(x_k)=H(t_k),
\end{equation}
one obtains 
\begin{equation}
T_{fi}(t)=i\epsilon _{n\bar{n}}<f\mid \sum_{k=1}^{\infty}T_{k}(t)
\int_{-t/2}^{t_{k}}dt_c\mid \!0\bar{n}_j\!>.
\end{equation}
Using the formula
\begin{equation}
\int_{-t/2}^{t/2}dt_1...\int_{-t/2}^{t_{k-1}}dt_k\int_{-t/2}^{t_k}dt_c
f(t_1,...,t_c)=\int_{-t/2}^{t/2}dt_c\int_{t_c}^{t/2}dt_1...\int_{t_c}^{t_{k-1}}
dt_kf(t_1,...,t_c),
\end{equation}
we change the integration order and pass on to the interval $(t,0)$. Finally
\begin{equation}
T_{fi}(t)=-\epsilon _{n\bar{n}}\int_{0}^{t}dt_c<\!f\!\mid T^{\bar{n}}(t-t_c)\mid\!
0\bar{n}_j\!>,
\end{equation}
\begin{eqnarray}
T^{\bar{n}}(t-t_c)=\sum_{k=1}T_k(t-t_c)=\nonumber\\
\sum_{k=1}^{\infty}(-i)^k \int_{t_c}
^{t}dt_1...\int_{t_c}^{t_{k-1}}dt_kH(t_1)...H(t_k).
\end{eqnarray}
Here $<\!f\!\mid T^{\bar{n}}(t-t_c)\mid\!0\bar{n}_j\!>$ is the matrix element of the 
antineutron annihilation in $\bar{n}$-nucleus in a time $\tau =t-t_c$. 

Equation (27) coincides with (64) of Ref. [9] except that in Eq. (64) of Ref. [9] the
matrix element $<\!f\!\mid T^{\bar{n}}(t-t_c)\mid\!0\bar{n}_p\!>$ corresponds to 
annihilation of $\bar{n}$ with the 4-momentum $p$ in the medium and not $\bar{n}$-nucleus.
For the problem under study this distinction is inessential. As in [8,9], the process (1)
probability $W(t)$ is found to be
\begin{equation}
W(t)\approx W_f(t)=\epsilon _{n\bar{n}}^2t^2,
\end{equation}
where $W_f(t)$ is the free-space $n\bar{n}$ transition probability. The result is precisely 
the same as for the $n\bar{n}$ transition in medium. The lower limit on the free-space 
$n\bar{n}$ oscillation time is $\tau ^b_{{\rm min}}=10^{16}$ yr. This value is interpreted as the estimation from above.

\section{Infrared divergence}
In this section it is shown that in the case of $S$-matrix problem formulation $(\infty,
-\infty)$ the amplitude of the model with bare ptopagator is singular: $T_{fi}(t\rightarrow 
\infty )\sim 1/0$. 

In Eq. (16) we put $t=\infty $. To realize the adiabatic hypothesis, we introduce the 
multiplier $exp (-\alpha \mid \!t_c\mid )$, $\alpha >0$. (In the previous section the 
adiabatic hypothesis has been not used since the limiting transition $t\rightarrow \infty $ 
was not made.) Then
\begin{eqnarray}
T_{fi}=-<f\mid \sum_{k=1}^{\infty}T_k(\infty )\int_{-\infty }^{t_k}dt_cH_{n\bar{n}}(t_c)
e^{-\alpha \mid \!t_c\mid }\mid \!0n_j\!>,\nonumber\\
T_k(\infty )=(-i)^k \int_{-\infty }^{\infty }dt_1...\int_{-\infty }^{t_{k-1}}dt_kH(t_1)...
H(t_k).
\end{eqnarray}
One obtains Eq. (19), where $t/2=\infty $ and
\begin{equation}
J_k(\infty )=\int d^3x_c\int_{-\infty }^{t_k}dt_c<T(\Psi _{\bar{n}}(x_k)
\bar{\Psi }_{\bar{n}}(x_c))>\bar{n}_j(x_c)e^{-\alpha \mid \!t_c\mid }.
\end{equation}
Since
\begin{eqnarray}
<T(\Psi _{\bar{n}}(x_k)\bar{\Psi }_{\bar{n}}(x_c))>=iG_{\bar{n}}(x_k,x_c)=\nonumber\\
-i\theta (t_k-t_c)\sum_{i}\bar{n}_i(x_k)\bar{n}_i^*(x_c)
\end{eqnarray}
(see (9)), $J_k(\infty )$ becomes
\begin{equation}
J_k=\int d^3x_c\int_{-\infty }^{t_k}dt_c\sum_{i}\bar{n}_i(x_k)\bar{n}_i^*(x_c)
\bar{n}_j(x_c)e^{-\alpha \mid \!t_c\mid }.
\end{equation}
In line with (6), (14) and (27) (see also (44)), $\bar{n}_m(t,{\bf x})=\exp (-i\epsilon _mt)\bar{n}_m({\bf x})$ and
\begin{equation}
J_k=\sum_{i}\int d^3x_c\int_{-\infty }^{t_k}dt_c\bar{n}_i({\bf x}_k)e^{-i\epsilon 
_it_k} \bar{n}_i^*({\bf x}_c)e^{i\epsilon _it_c}\bar{n}_j({\bf x}_c)e^{-i\epsilon _jt_c} 
e^{-\alpha \mid \!t_c\mid }.
\end{equation}
Taking into account that
\begin{equation}
\int_{-\infty }^{t_k}dt_ce^{i(\epsilon _i-\epsilon _j)t_c-\alpha \mid \!t_c\mid }=
\frac{1}{i}\frac{e^{i(\epsilon _i-\epsilon _j)t_k}}{\epsilon _i-\epsilon _j-i\alpha },
\end{equation}
we get
\begin{eqnarray}
J_k=\sum_{i}\frac{1}{i}\frac{e^{-i\epsilon _jt_k}}{\epsilon _i-\epsilon _j-i\alpha }
\bar{n}_i({\bf x}_k) \int d^3x_c\bar{n}_i^*({\bf x}_c)\bar{n}_j({\bf x}_c)=\nonumber\\
\frac{1}{i}\frac{e^{-i\epsilon _jt_k}}{\epsilon _j-\epsilon _j-i\alpha }\bar{n}_j({\bf x}_k)
\sim \frac{\bar{n}_j(x_k)}{0}.
\end{eqnarray}
As in the case of $n\bar{n}$ transitions in medium (see Eqs. (53)-(55) of Ref. [9]), the 
amplitude diverges. The value $1/(\epsilon _j-\epsilon _j)$ plays the rule of singular 
propagator.

\section{Model with dressed propagator}
In the model considered above the matrix element  $<\!f\!\mid T^{\bar{n}}(t-t_c)\mid\!0
\bar{n}_j\!>$ (see (27)) and amplitude $M_a$ involve all the $\bar{n}$-nuclear interactions followed by annihilation including the antineutron rescattering in the initial state. In principle, the part of this interaction can be included in the antineutron Green function [9,10,17]. Then the antineutron self-energy $\Sigma $ is generated. In Eq. (36) one should replace
\begin{equation}
\frac{1}{\epsilon _j-\epsilon _j-i\alpha }\rightarrow  \frac{1}{\epsilon _j-\epsilon _j
-\Sigma -i\alpha }\neq \frac{1}{0}.
\end{equation}
In this case the amplitude is non-singular and calculation is standard. We consider the
process (2) for simplicity. As in [10,17], the process probability is found to be 
\begin{equation}
W_d(t)\approx \frac{\epsilon _{n\bar{n}}^2}{\Sigma ^2}\Gamma t
\end{equation}
($\Sigma $ is the parameter), whereas the potential model gives the inverse $\Gamma $-dependence [4-11] 
\begin{equation}
W_{{\rm pot}}(t)\approx \frac{4\epsilon _{n\bar{n}}^2t}{\Gamma }.
\end{equation}
The calculations in the framework of unitary model tend to increase the $n\bar{n}$ transition probability: 
\begin{equation}
\frac{W_d}{W_{{\rm pot}}}=\frac{\Gamma ^2}{4\Sigma ^2}>1.
\end{equation}
The lower limit increases as well. Let $\tau ^d_{{\rm min}}$ and $\tau _{{\rm pot}}$ be the
lower limits on the free-space $n\bar{n}$ oscillation time obtained by means of Eqs. (38) and 
(39), respectively. It is easy to verify that [10]
\begin{equation}
\tau ^d_{{\rm min}}=\frac{\Gamma }{2\Sigma }\tau _{{\rm pot}}.
\end{equation}
If $\Gamma =100$ MeV and $\Sigma =10$ MeV then 
\begin{equation}
\tau ^d_{{\rm min}}=5\tau _{{\rm pot}}.
\end{equation}
For $\tau _{{\rm pot}}=2.36\cdot 10^{8}$ s [7] Eq. (42) gives
\begin{equation}
\tau ^d_{{\rm min}}=1.2\cdot 10^{9}\; {\rm s},
\end{equation}
which exceeds the lower limit given by the Grenoble reactor experiment [3] by a factor of 14. The parameter $\Sigma $ is uncertain. We have put $\Sigma ={\rm Re}U_{\bar{n}}-U_n\approx 10$ MeV only for estimation ($U_n$ and $U_{\bar{n}}$ are the potentials of neutron and antineutron). If $\Sigma =0$, we come to the model with bare propagator. 

If $\Sigma \rightarrow 0$, $W_d(t)$ rises quadratically. So $\tau ^d_{{\rm min}}$ is 
interpreted as the estimation from below (conservative limit). 

\section{Summary and conclusion}
Result (38) corresponds to the model with non-singular amplitude. Although there is no infrared 
singularity, this model has {\em essential drawbacks}. The model as well as possible suppression 
mechanisms are studied in [9,17]. In present paper the particular attention was given to the model 
with bare propagator since the corresponding calculations are non-typical. 

It is significant that $W_d(t)$ rises quadratically as $\Sigma \rightarrow 0$. This circumstance should be 
clarified; otherwise the model under study can be rejected. The calculation in the framework 
of the model with bare propagator gives the finite result, which justifies our approach from a conceptual 
point of view and consideration of the model with bare propagator at least as the limiting case. In fact 
this model seems quite realistic in itself [9,17]. In this connection we recall the reasons owing to which 
the approach with finite time interval has been used.

Since the $S$-matrix should be unitary, the calculation should be done beyond the potential
model. However, the $S$-matrix amplitude based on hermitian Hamiltonian contains unremovable peculiarity. Moreover, for the problem under study the $S$-matrix problem formulation $(\infty ,-\infty )$ is physically incorrect [9]. For these reasons the problem is considered on the interval $(t,0)$.

On the other hand, if the problem is formulated on the finite time interval, the decay 
width $\Gamma $ cannot be introduced since $\Gamma =\sum_{f\neq i}\mid S_{fi}(\infty,
-\infty)\mid ^2/T_0$, $T_0\rightarrow \infty $. This means that the standard calculation 
scheme should be completely revised.

We point to the important detail of the model which explains the absence of the suppression. It is seen from Eqs. (14) and (27) that due to zero momentum transfer both pre- and post-$n\bar{n}$ conversion spatial wave functions of the system coincide:
\begin{equation}
\mid\!0\bar{n}_j\!>_{sp}=\mid\!0n_j\!>_{sp}.
\end{equation}
Recall that the neutron potential is included in $H_0$; the $\bar{n}$-nuclear interaction
$H$ is involved in $H_I$. Since the Hamiltonian $H$ acts on the antineutron, it is turned 
on following the forming of the $\bar{n}$-nucleus and so Eq. (44) takes place. There is no 
energy gap $U_{\bar{n}}-U$ which can lead to a very strong suppression of $n\bar{n}$ 
transition. The Hamiltonian of the $n\bar{n}$ transition changes only the internal quantum
numbers of neutron: $H_{n\bar{n}}\mid \!0n_j\!>=\epsilon _{n\bar{n}}\mid \!0\bar{n}
_j\!>$ [4].

Relation (44) explains the result (29). Nevertheless, we view the results of the model 
with bare propagator with certain caution since the process is extremely sensitive to the 
details of the model. Indeed, in the models with bare and dressed propagators the interaction 
Hamiltonians $H_I$ and unperturbed Hamiltonians are the same. The sole physical distinction 
between models is the zero antineutron self-energy in the model with bare 
propagator. However, it leads to the fundamentally different results (see (29) and (38)). 
This is because the amplitude (12) is in the peculiar point. Due to this the problem is 
extremely sensitive to the value of antineutron self-energy $\Sigma $ as well as the 
description of initial neutron state [18] and the value of momentum transfered in the 
$n\bar{n}$ transitions vertex [17]. 

In conclusion, new model of $n\bar{n}$ transitions in nuclei based on unitary $S$-matrix has
been considered. Since the results are the same as for nuclear matter, the conclusions are 
identical as well (see Sect. 5 of Ref. [17]): taking into account the result sensitivity to 
the details of the model, the values $\tau ^d_{{\rm min}}=1.2\cdot 10^{9}$ s and 
$\tau ^b_{{\rm min}}=10^{16}$ yr are interpreted as the estimations from below (conservative 
limit) and from above, respectively. So the realistic limit $\tau _{{\rm min}}$ can be in the range
\begin{equation}
10^{16}\; {\rm yr}>\tau _{{\rm min}}>1.2\cdot 10^{9}\; {\rm s}.
\end{equation}

The estimation from below $\tau _{{\rm min}}>1.2\cdot 10^{9}\; {\rm s}$ exceeds the restriction
given by potential model by a factor of 5 and the lower limit given by the Grenoble reactor experiment 
[3] by a factor of 14. At the same time the range of uncertainty of $\tau _{{\rm min}}$ is too wide. 
Further investigations are desirable.

\end{document}